\documentclass[11pt,a4paper]{article}
\usepackage{delarray,amsfonts,amsmath,amssymb,graphicx}
\DeclareGraphicsExtensions{ps,eps,ps.gz} \headheight 0.6cm
\def\@citex[#1]#2{\if@filesw\immediate\write\@auxout
        {\string\citation{#2}}\fi
\def\@citea{}\@cite{\@for\@citeb:=#2\do
        {\@citea\def\@citea{,}\@ifundefined
        {b@\@citeb}{{\bf ?}\@warning
        {Citation `\@citeb' on page \thepage \space undefined}}
        {\csname b@\@citeb\endcsname}}}{#1}}
\newif\if@cghi
\def\cite{\@cghitrue\@ifnextchar [{\@tempswatrue
        \@citex}{\@tempswafalse\@citex[]}}
\def\citelow{\@cghifalse\@ifnextchar
[{\@tempswatrue\@citex}{\@tempswafalse\@citex[]}}
\def\@cite#1#2{{\if@cghi\unskip$\null^{#1}$\else #1\fi\if@tempswa\typeout
        {warning: optional citation argument ignored: `#2'} \fi}}

\renewcommand{\thefootnote}{\alph{footnote}}
\def\tt{{\mathbb T}_\theta}

\def\aa{{\cal A}}
\def\hh{{\cal H}}
\def\rr{{\mathbb{R}}}
\def\cc{{\mathbb C}}

\newcommand{\abs}[1]{\lvert#1\rvert}

\newcommand{\scl}[2]{\langle#1,#2\rangle}

\def\dm{\lp\begin{array}}
\def\fm{\end{array}\rp}
\def\dbb{\lb\begin{array}}
\def\fbb{\end{array}\rb}
\def\dbn{\left.\begin{array}}
\def\fbn{\end{array}\right.}

\def\lb{\left[}
\def\rb{\right]}
\def\lp{\left(}
\def\rp{\right)}

\def\ms{mod\`ele standard}

\def\m3{M_3 \lp \cc \rp}
\def\m2{M_2 \lp \cc \rp}

\def\cc{{\mathbb{C}}}
\def\rr{{\mathbb{R}}}

\def\aa{{\cal A}}

\def\hh{{\cal H}}

\def\oo{{\cal O}}

\def\mm{{M}}

\def\L2{L_2(\mm)}

\def\xo0{\omega^0_x}
\def\yo0{\omega^0_y}

\def\o0{\omega_0}

\def\xo0{x_\omega^0}
\def\yo0{y_\omega^0}

\def\tt{\mathcal{T}}

\begin{document}
\title{Is life a thermal horizon ?}
\author{Pierre Martinetti,\\ Universit\`a di Roma "La Sapienza",\\
5 piazzale Aldo Moro, 00185 Roma, Italia\\{\it
pierre.martinetti@laposte.net}}  \maketitle
\date
\begin{abstract}
This talk aims at questioning the vanishing of Unruh temperature for
an inertial observer in Minkovski spacetime with finite lifetime,
arguing that in the non eternal case the existence of a causal
horizon is not linked to the non-vanishing of the acceleration. This
is illustrated by a previous result, the diamonds temperature, that
adapts the algebraic approach of Unruh effect to the finite
case.{\it Written for the proceedings of DICE 2006, Piombino
september 2006.}

\end{abstract}

\section{Introduction}
For a uniformly accelerated observer with {\it infinite lifetime},
the vacuum state of a suitable quantum field theory in Minkovski
spacetime $M$ appears as a thermal equilibrium state with
temperature
\begin{equation}
\label{effetunruh}
 T_U = \frac{\hbar a}{2\pi k_b c}.
\end{equation}
This result, known as the Unruh effect, can be derived in at least
three ways: a comparison of quantization schemes on various regions
of $M$, an integration along the worldline of the observer of the
interaction term between a detector and the vacuum, or a geometrical
approach based on the modular properties of the region causally
accessible to the observer. Those three approaches strongly rely on
the eternity of the observer and in this talk we would like to
discuss what is to be expected for a non-eternal observer. In
particular, at the light of the analogy between Unruh and Hawking
temperatures, it appears that for an inertial observer (i.e. with
zero acceleration) with {\it finite lifetime} there is no clear
reason to ask for the vanishing of the thermal effect. The reason is
that for a non-eternal observer the presence of an horizon is not
linked to the acceleration as in the eternal case. This point is
discussed in section 2. In section 3 we give a presentation of the
geometric approach to Unruh effect, based on the KMS formulation of
statistical physics, the modular theory of Tomita-Takesaki and the
algebraic formulation of quantum field theory. We then recall how to
use these technics to treat the non-eternal case\cite{diamonds},
yielding an "Unruh-effect for bounded trajectories" that has the
striking property of being non-zero for zero acceleration.

 This talk
aims at underlining that the non-vanishing of the temperature for a
finite lifetime inertial observer is not unexpected as it might seem
at first sight, but is rather natural. In fact, and this will be our
conclusion in section 4, if the thermal properties of the vacuum had
to disappear for an inertial observer with finite lifetime, this
would raise the following question (whose answer to our knowledge is
not clear): what makes the horizon of a finite lifetime observer -
its "life horizon" given by the intersection of the future cone of
its birth with the past of its death - so different from the horizon
of an eternal observer - a Rindler wedge - so that to kill the
thermal property of the vacuum ?

\section{The importance of being eternal}

Let us recall three main ways to Unruh effect:

- In Unruh's original approach\cite{unruh}, $T_U$ is obtained by
observing that the vacuum for a quantization scheme on all $M$ is
not a pure state for an alternative (but as well defined)
quantization prescription on the Rindler wedge $W$\footnote{in
cartesian coordinates $W$ is the set of points such that $x>\abs{t}$
where $x$ is the direction of the acceleration}.

- In DeWitt approach\cite{dewitt} one integrates all along the
worldline of a uniformly accelerated and eternal oberver the
interaction term of a quantum system coupled to the vacuum. And it
turns out that the system gets excited as if it was at rest in a
thermal bath at temperature $T_U$.

- Finally one can recover $T_U$, independently of any detector
prescription, by noting that the trajectory of a uniformly eternal
accelerated observer coincides with the orbit of the point at
coordinate $t=0$ under the action of the vacuum modular group
associated to the algebra of observables on the wedge region. By
general result of modular theory, one knows that the vacuum is a
thermal state at temperature $\beta^{-1}$ with respect to the time
evolution determined by the modular group, up to a rescaling
$$\tau = -\beta s$$
where $\tau$ denotes the physical time and $s$ the modular
parameter. In case of the wedge $W$, the comparison of the two
parameterizations of the trajectory, one by $s$ the other one by
$\tau$, precisely yield $\beta = T_U^{-1}$.

Note that in those three approaches the eternity of the observer is
an important requirement: either one needs to integrate the
interaction term from $-\infty$ to $+\infty$ in order to recover a
thermal distribution, or one uses some property of the wedge $W$.
The latest is physically relevant for it is the (whole and only)
region of $M$ with whom an eternal uniformly accelerated observer
can interact. For a non eternal observer, $W$ is no longer
significant and considering a quantization scheme on $W$ or the
modular group of $W$ has no more physical meaning.

Eternity of the observer is generally overcame by viewing $T_U$ as a
limit for asymptotic states. However such a limit is not always
meaningful. Specifically $T_U$ identifies to the Hawking temperature
measured by an observer very close to the horizon of an eternal
black hole (see [\citelow{birrell}] for instance) but not for a Kerr
black hole. Quoting Wald\cite{wald}: \noindent "{\it{ the difference
in nature between the Unruh effect [...] and the Hamking effect of
particle creation by black holes [...] is dramatically illustrated
by considering the case of the Kerr metric [...] In essence this is
because there is {\underline{no analog of incoming thermal radiation
from infinity} with respect to the notion of time translations
defined by the Killing field which generates the horizon, since this
Killing field has spacelike orbits near infinity. However there is
no corresponding difficulty with the derivation of particle creation
in the case where gravitational collapse produces a Kerr black
hole.}}} In this framework the question of an Unruh effect for a non
eternal observer becomes relevant independently of the asymptotic
acceptation. A traditional answer is to consider that, at best, the
excitation rate at finite time might give an indication on the
duration of the interaction time between the detector and the
thermal bath. However this answer raises more questions than it
solves:

-first it makes no distinction between the fact that the vacuum seen
from the accelerated point of view is a thermal state, and the fact
that a detector may need an infinite amount of time to get in
equilibrium with it.

-second it questions the origin of the effect. If, as suggested by
the analogy with Hawking temperature, $T_U$ emerges due to the
presence of an horizon, then a finite lifetime observer should also
see the vacuum as a thermal state. Indeed the double cone region (or
diamond region)
$$D = \text{birth}^+ \cap \text{death}^-$$
where $\pm$ indicates future/past cones, acts for a non eternal
observer as the wedge for an eternal one (see figure 1). Unless one
is able to point out a difference in nature between $D$ and $W$ (is
compactness such important in this framework ?), one should be ready
to accept that the vacuum is thermal also from a non-eternal
accelerated point of view.

Assuming this point is not academic. In the eternal case the
presence of an horizon coincides with the non vanishing of the
acceleration: an inertial, eternal observer has
 access to the whole of Minkovki spacetime. There is no longer wedge to
 limit its horizon and $T_U$ vanishes. On the contrary in the finite lifetime case the two notions,
 horizon and acceleration, do not coincide. For
 such an observer $D$  acts like a causal horizon, whether he is accelerated or not.
\begin{figure}

 \begin{center}
\hspace{0truecm}\mbox{\rotatebox{0}{\scalebox{.5}{\includegraphics{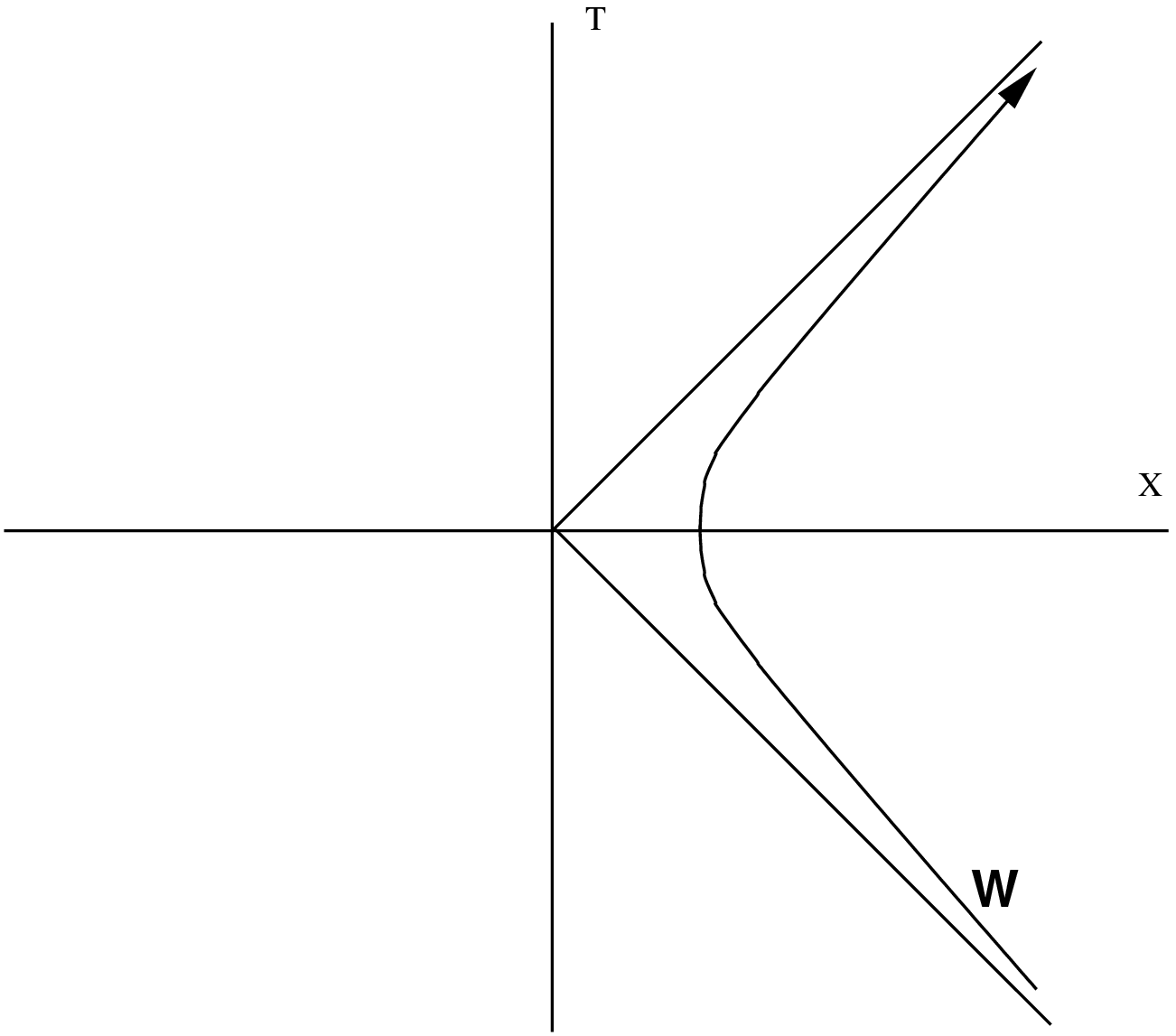}\hspace{6truecm}\includegraphics{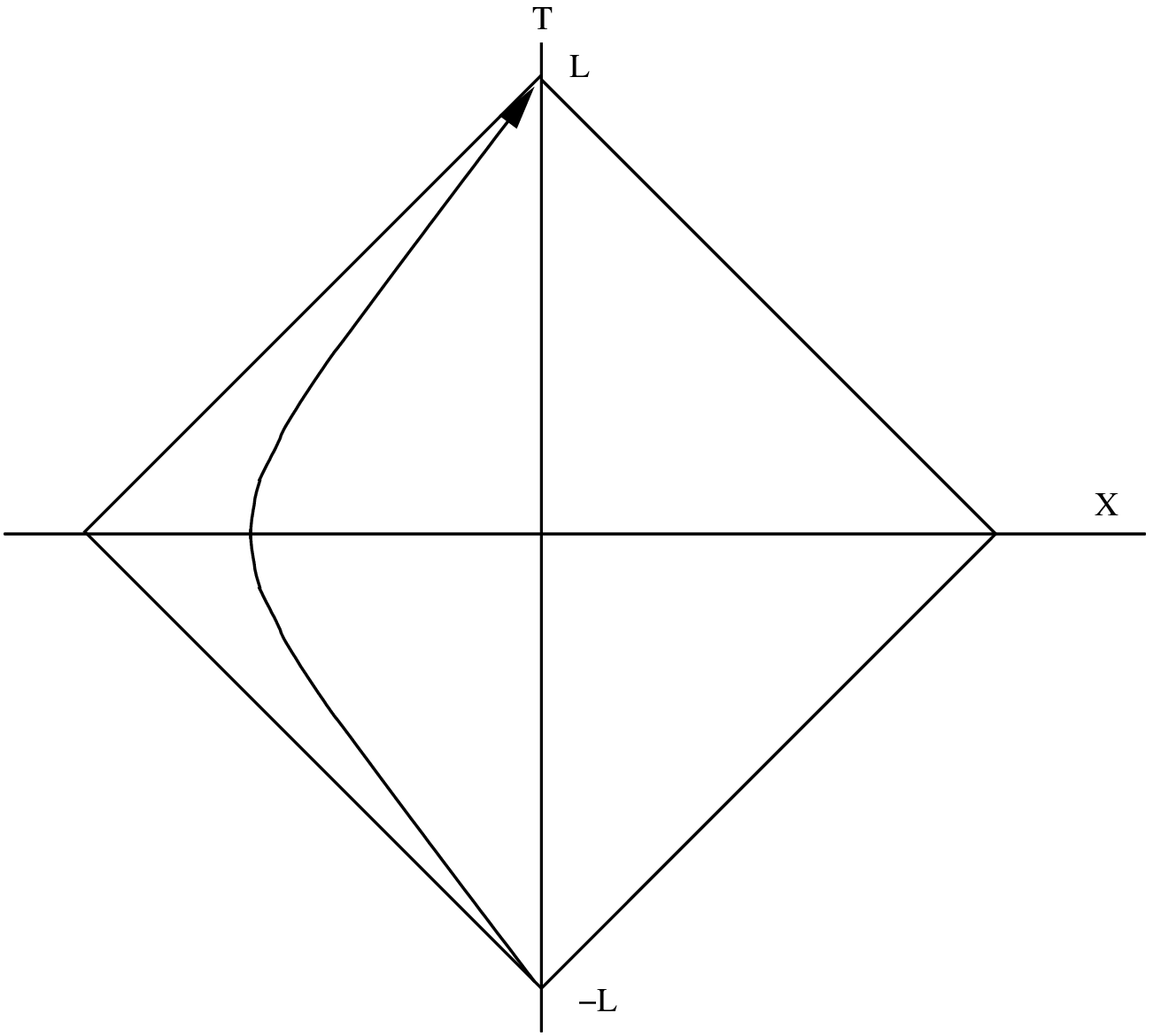}}}}
\caption{\hspace{0truecm} Rindler Wedge\hspace{3truecm} Double
wedge}\end{center}
\end{figure}
Hence there is no reason why an inertial observer with lifetime
$\mathcal{T}$ should not observe an Unruh temperature
$T_U(\mathcal{T})\neq 0$. The only requirement to be compatible with
the eternal case is $\underset{\mathcal{T}\rightarrow
+\infty}{\lim}T_U(\mathcal{T})=0$.

\section{The algebraic way and the non eternal case}

Motivated by completely different reasons (namely, the issue of time
in quantum gravity\footnote{which in the present context can be
stated as follows: assuming that covariance is preserved at the
quantum level and that one is surrounded by a quantum superposition
of states of the gravitational field, then one can a priori picks
out any direction as {\it the} direction of time. How to combine
this freedom at the quantum level with the locally unique intuition
of physical flow of time at the classical level ?} and a possible
solution known as {\it the thermal time hypothesis}
\cite{carloconnes}) we proposed in [\citelow{diamonds}] an
adaptation for Unruh effect in Minkovski spacetime for an observer
with finite lifetime $\mathcal{T}$. Our construction is based on the
algebraic approach \cite{bisognano,sewell} that we recall below.

Given a statistical system with algebra of observables $\aa$ and
Hamiltonian $H$, a state $\omega$ is said to be KMS with parameter
$\beta$ if it satisfies
\begin{equation}
\label{kms} \omega(\alpha_t(a)b) =
\omega(b\alpha_{t+i\beta}(a))\quad\quad a,b\in\aa
\end{equation}
where $\alpha_t(a)= e^{iHt}ae^{-iHt}$ is the time translation,
extended to complex variables\footnote{as well, one asks that the
function $z\mapsto \omega(b\alpha_z(a))$ be analytic in the strip
$0<\text{Im}\, z<\beta$ for any $a,b\in\aa$}. It has been shown (see
 [\citelow{haag}] for an complete presentation of the subject) that for a
system with a finite number of degrees of freedom, being KMS with
parameter $\beta$ is equivalent to being a Gibbs equilibrium state
at temperature $\beta^{-1}$. Moreover contrary to Gibbs definition
KMS properties are still meaningful at the thermodynamical limit.
Therefore given a system with a well known time evolution $\alpha_t$
one defines an equilibrium state in the following way:

\begin{center}
\framebox{\parbox{15.5cm}{An equilibrium state at temperature
$\beta^{-1}$ is a state that satisfies the KMS condition -with
parameter $\beta$- with respect to the time evolution $\alpha_t$.}}
\end{center}

\noindent In other terms, given a temperature and a time evolution,
the KMS condition allows to characterize thermal equilibrium states,
\begin{equation}
\label{dessin} \left\{\begin{array}{l} \text{time} \\
\text{temperature}\end{array} \longrightarrow \text{thermal
state.}\right.
\end{equation}
Now it happens that mathematics furnish a lot of KMS state.
Specifically to any Von Neumann algebra $\aa$ acting on an Hilbert
space $\hh$ is associated a canonical (up to unitary) $1$-parameter
group of automorphisms\cite{tomtak}
\begin{equation}
s\in\rr \mapsto\sigma_s\in\text{Aut}(\aa)
\end{equation}
 built from a cyclic and separating
vector $\Omega\in\hh$. Explicitly the {\it modular group} $\sigma_s$
is
\begin{equation}
\sigma_s(a)= \Delta^{is} a \Delta^{-is}\quad\quad a\in\aa,\;
s\in\mathbb{R}
\end{equation}
where $\Delta$ is given by the polar decomposition of Tomita's
operator $S$ defined on $\aa\Omega\subset\hh$ by $ Sa\Omega =
a^*\Omega. $ The remarkable point is that the state defined by
$\Omega$ is KMS with respect to $\sigma_s$
\begin{equation}
\scl{\Omega}{\sigma _s(a)b\Omega} = \scl{\Omega}{b\sigma
_{s_-i}(a)\Omega}.
\end{equation}
Putting $\alpha_s \doteq \sigma_{-\beta s}$ where $\beta$ is a fixed
constant, one is back to (\ref{kms}) as soon as
\begin{equation}
\label{rescale} t= -\beta s.
\end{equation}
In other terms, as a reformulation of the KMS-definition, one
has\cite{haag}:

 \begin{center}
\framebox{\parbox{16.5cm}{A thermal state at temperature
$\beta^{-1}$ is a state whose associated modular group $\sigma_s$
coincides with the time flow $\alpha_t$ up to rescaling
(\ref{rescale}).}}
\end{center}

\noindent This definition is less tractable than Gibbs's one. But it
has the advantage to give one solution to the issue of time, simply
by inverting the arrow of (\ref{dessin})
\begin{equation}
\label{dessin2} \left\{\begin{array}{l} \text{thermal state} \\
\text{temperature}\end{array} \longrightarrow \text{time.}\right.
\end{equation}
Namely assuming that time flow is not known a priori but the system
is in an equilibrium state $\omega$ at temperature $\beta^{-1}$, the
thermal time hypothesis maintains that the flow of time is given by
the modular flow associated to $\omega$, the physical time $t$ being
related to the modular parameter $s$ by (\ref{rescale}). For physic,
the difficulty is of course to explicitly compute the modular flow.
This has been done for the wedge region by Bisognano and
Wichman\cite{bisognano}: taking for $\aa$ the algebra of local
observables{\footnote{obtained by smearing out on $W$ the fields
viewed as operator valued distributions\cite{haag,wightman}} on the
wedge $W$ and for $\Omega$ the vacuum state, one has that the
modular group coincides with the time flow of a uniformly
accelerated observer $\oo$. So the line of universe of $\oo$ can be
parameterized by its proper time $\tau$ or by the modular parameter
$s$ and the ratio of the corresponding tangent vectors precisely
yield Unruh temperature
\begin{equation}
\label{temp} \frac{ds}{d\tau} = \text{ constant} = -T_U.
\end{equation}

A natural question is whether the same analysis,
\begin{equation}
\label{dessin2} \left\{\begin{array}{l} \text{thermal state} \\
\text{time}\end{array} \longrightarrow \text{temperature,}\right.
\end{equation}
 is true for other
regions of Minkowski spacetime. In [\citelow{diamonds}] we have
considered the modular group associated to the region causally
connected to a uniformly accelerated observer $\mathcal{O}$ with
lifetime\footnote{the observer's proper time $\tau$ is measured from
$-\tau_0$ to $\tau_0$.}
${\mathcal T} = 2\tau_0,$
namely a diamond-shape region $D$
$$\abs{\overrightarrow{x}} +
\abs{t} < L(a,\tau_0)
$$
where
$$L(a,\tau_0) \doteq
a^{-1}\text{argsinh} a\tau_0$$ is the size of the diamond, and
coincides with (half of) the lifetime of the observer in case of
zero acceleration
$$2L(0,\tau_0)= 2\tau_0 = {\mathcal{T}}.$$
Assuming the field is conformally invariant, Hislop and
Longo\cite{hislop} have shown that the line of universe of
$\mathcal{O}$ is nothing but the orbit of one of its point under the
action of the modular group. In other terms, as in the eternal case,
the trajectory of $\mathcal{O}$ can be parameterized either by
$\mathcal{O}$'s proper time $\tau\in]-\tau_0,\tau_0[$ or by the
modular parameter $s\in]-\infty, +\infty[$. But now the ratio of the
two parameterizations is no longer constant since the proper time
$\tau$ is bounded while the modular parameter $s$ is unbounded.
Explicit computations detailed in [\citelow{diamonds}] yield
\begin{equation}
\label{tempe} \frac{ds}{d\tau} = T(\tau_0, \tau) = T_U \frac{\sinh
a\tau_0}{\cosh a\tau_0 - \cosh a\tau}.
\end{equation}
We have interpreted (\ref{tempe}) as a temperature by noting that
for given $\tau_0$ and acceleration $a$, $T(\tau_0, \tau)$ is almost
a constant for most of $\mathcal{O}$'s lifetime and takes the value
observed in the middle of its life,
$\label{temp}
 T(\tau_0, \tau) \backsimeq T(\tau_0, 0)$ or,
written as a function of $\tau_0$ and $a$,
\begin{equation}
\label{diamtemp} T(a, \tau_0) \doteq T_U \frac{\cosh a\tau_0 +
1}{\sinh a\tau_0}.
\end{equation}
The interesting point is that this temperature does not vanish for
an inertial observer
\begin{equation}
\label{diamtemp} T(0, \tau_0) = \frac{2}{\pi{\mathcal T}}
\end{equation}
as soon as the lifetime ${\mathcal T}$ is finite.

\section{Conclusion}

Several adaptations of Unruh effect have been proposed for an
observer with a finite lifetime (see
 [\citelow{diamonds,padmanabhan,schlicht,casadio,louko}] for the most
recent). By this one often means that the detector does interact
with the vacuum only for a finite period of time. The result
generally depends on the nature of the coupling with the vacuum as
well as on the shape of its switching on/off (as explained by J.
Louko in his talk). What seems to be commonly admit is that the
vacuum is still thermal at temperature $T_U$ but the detector has no
time to reach the thermal equilibrium. Here we argued that the
temperature of the vacuum as seen from a {\it finite lifetime}
accelerated point of view:

- does not necessarily equal $T_U$ and can be given by some
corrections in ${\mathcal T}^{-1}$,

- has no reason to vanish for an inertial observer since, in the
finite case, the presence of a diamond shape "life horizon" does not
depend on the acceleration.

The thermal time hypothesis allows to derive a temperature for this
"life horizon", whose interpretation is questionable:  (\ref{tempe})
is not a constant with respect to the proper time of the observer,
but a function which is almost constant on most of its domain.
Basically what the thermal time hypothesis allows is to conformally
map the infinity of the lifetime to a (sharp) infinity of the
temperature. What is best for physical interpretation is not clear
to us at the moment. However, independently of this specific
proposal, it remains that if life is not a thermal horizon for an
inertial observer, then one should explain why $W$ leads to a
thermalization of the vacuum whereas $D$ does not. The only obvious
difference is that $D$ is compact while $W$ is not. So the vanishing
of $T_U$ for inertial observer, eternal or not, would imply that
compactness of the horizon has something to do with Unruh
temperature. This could be interesting to confront this idea to
general study on horizons\cite{jacobparent}. Otherwise it would be
interesting to study whether the thermalization of the vacuum for
inertial observer leads to some contradiction with known physics.
\\

\noindent{\bf Acknowledgments} work supported by a Marie Curie
fellowship EIF-025947-QGNC.
\\

\end{document}